\documentclass[aps,letterpaper,twocolumn,preprintnumbers,floatfix,superscriptaddress,nofootinbib]{revtex4-1}

\usepackage{graphicx}  % needed for figures
\usepackage{dcolumn}   % needed for some tables
\usepackage{bm}        % for math
\usepackage{amssymb, amsmath}
\usepackage{slashed}   % for math
\usepackage{color}
\usepackage[bookmarks]{hyperref}

\usepackage{tensor} % for dealing with the position of indexes in tensors

\usepackage[usenames,dvipsnames,svgnames,table]{xcolor}

\def\beq{\begin{equation}}
\def\eeq{\end{equation}}
\def\M{\mathcal{M}}  %amplitude%
\def\mp{m_{\mathrm{Pl}}} %reduced planck mass%
\def\E{\mathcal{E}}  %units for energy%

\begin{document}

\widetext
\leftline{Saclay-t17/151}
\leftline{CERN-TH-2017-201}
\leftline{SISSA 46/2017/FISI}

\title{Beyond Amplitudes' Positivity and the Fate of Massive Gravity}
\author{Brando Bellazzini}
\affiliation{Institut de Physique Th\'eorique, Universit\'e Paris Saclay, CEA, CNRS, F-91191 Gif-sur-Yvette, France}
\affiliation{Dipartimento di Fisica e Astronomia, Universit\`a di Padova, Via Marzolo 8, I-35131 Padova, Italy}
\author{Francesco Riva}
\affiliation{Theory Division, CERN, CH-1211 Geneva 23, Switzerland}
\author{Javi Serra}
\affiliation{Theory Division, CERN, CH-1211 Geneva 23, Switzerland}
\author{Francesco Sgarlata}
\affiliation{SISSA International School for Advanced Studies and INFN Trieste, via Bonomea 265, 34136, Trieste, Italy}

%\date{\today}

\begin{abstract}
\noindent 

We constrain effective field theories  by going beyond the familiar positivity bounds that follow from unitarity, analyticity, and crossing symmetry of the scattering amplitudes.  
As interesting examples, we discuss the implications of the bounds for the Galileon and  ghost-free massive gravity.    
The combination of our theoretical bounds with the experimental constraints on the graviton mass implies that the latter is either ruled out or unable to describe gravitational phenomena, let alone to consistently implement the Vainshtein mechanism, down to the relevant scales of fifth-force experiments, where general relativity has been successfully tested.
We also show that the Galileon theory must contain symmetry-breaking terms that are at most one-loop suppressed compared to the symmetry-preserving ones.
We comment as well on other interesting applications of our bounds.

\end{abstract}

%\pacs{yyy}

\maketitle

%%%%%%%%%%%%%%%%%%%%%%%%%%%%%%%%%
\section{Introduction and summary}

The idea that physics at low energy can be described in terms of light degrees of freedom alone is one of the most satisfactory organising principle in physics, which goes under the name of Effective Field Theory (EFT). A quantum field theory (QFT) can be viewed as the trajectory in the renormalization group flow from one EFT to another, each  being  well described by an approximate fixed point where the local operators are classified mainly by their scaling dimension. The effect of ultraviolet (UV) dynamics is systematically accounted for in the resulting infrared (IR) EFT by integrating out the heavy degrees of freedom, which generate an effective Lagrangian made of infinitely many local operators. Yet, EFT's are predictive even when the UV dynamics is unknown, because in practice only a finite number of operators  contributes, at a given accuracy, to observable quantities. The higher the operator dimension, the smaller the effect at low energy. 

Remarkably, extra information about the UV can always be extracted if the underlying Lorentz invariant microscopic theory is unitary, causal and local.
These principles are stirred in the fundamental properties of the S-matrix such as unitarity, analyticity, crossing symmetry, and polynomial boundedness. These imply a UV-IR connection in the form of dispersion relations that link the (forward) amplitudes in the deep IR with the discontinuity across the branch cuts integrated all the way to infinite energy \cite{GellMann:1954db,Goldberger:1955zz}. Unitarity ensures  the positivity of such discontinuities, and in turn the positivity of (certain) Wilson coefficients associated to the operators in the IR effective Lagrangian. This UV-IR connection can be used to show that Wilson coefficients with the ``wrong'' sign can not be generated by a Lorentz invariant, unitary, casual and local UV completion, as it was emphasised e.g.~in Ref.~\cite{Adams:2006sv}. These positivity bounds have found several applications, including the proof of the a-theorem \cite{Komargodski:2011vj,Luty:2012ww}, the study of chiral perturbation theory~\cite{Manohar:2008tc}, $WW$-scattering and theories of composite Higgs \cite{Distler:2006if,Vecchi:2007na,Bellazzini:2014waa,Low:2009di,Falkowski:2012vh,Urbano:2013aoa}, as well as quantum gravity \cite{Bellazzini:2015cra}, massive gravity \cite{Cheung:2016yqr,Bonifacio:2016wcb,Bellazzini:2016xrt}, Galileons \cite{Nicolis:2009qm,Keltner:2015xda,Bellazzini:2016xrt,deRham:2017imi}, inflation \cite{Baumann:2015nta,Croon:2015fza}, the weak gravity conjecture \cite{Cheung:2014ega,Cheung:2014vva} and conformal field theory \cite{Komargodski:2016gci,Hartman:2015lfa,Alday:2016htq}. The approach has been recently extended  to particles of arbitrary spin~\cite{Bellazzini:2016xrt}, with applications to massive gravity and the EFT of a Goldstino \cite{Bruggisser:2016ixa,Bellazzini:2017neg,Bellazzini:2017bkb}, and it has led to the formulation of a general no-go theorem  on the leading energy-scaling behavior of the amplitudes in the IR~\cite{Bellazzini:2016xrt}. Ref.'s~\cite{deRham:2017avq,deRham:2017imi,deRham:2017zjm} extended this technique beyond the forward limit, providing an infinite series of positivity constraints for amplitudes of arbitrary spin.\\

%
%%%%%%%%%%
\begin{figure}[ht]
\centering
\includegraphics[width=8.7cm]{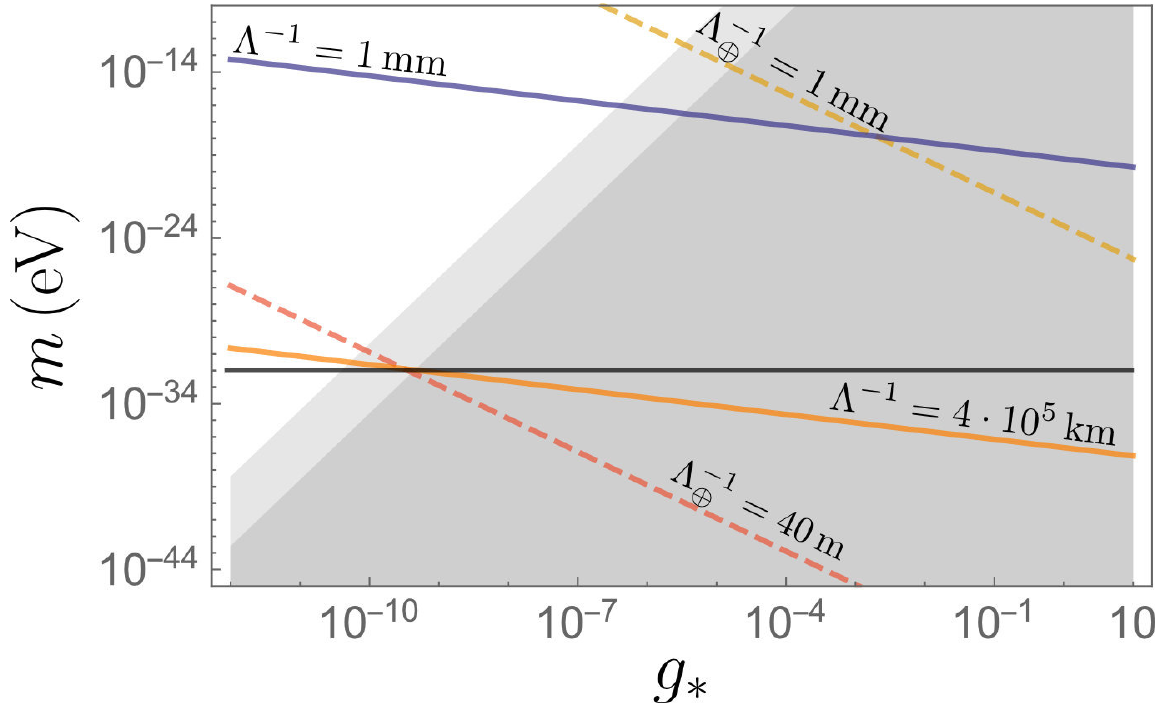}
\caption{Exclusion region for massive gravity in the plane of $(g_*,m)$,  where $g_*= (\Lambda/\Lambda_3)^3$  is the hierarchy between the physical cutoff $\Lambda$ and the strong coupling scale $\Lambda_3$, and $m$ is the graviton mass. The gray region is theoretically excluded by our lower bound Eq.~(\ref{FVSmaxrelation}), with accuracy either $\delta=1\%$ (dark) or $\delta=5\%$ (light), irrespectively of the values for $(c_3,d_5)$ in the massive graviton potential. Colored lines show the physical cutoff length: solid lines correspond to $\Lambda$ in Eq.~(\ref{barecutoff}), while dashed lines correspond to $\Lambda_\oplus$, obtained after assuming ad-hoc a Vainshtein redressing of $\Lambda$ due to the gravitational field on the Earth's surface, Eq.~(\ref{VainCutoff}). 
Either cutoff, and with it the domain of predictivity of massive gravity, increases with $g_*$ and $m$, at odds with our theoretical constraint and the experimental upper bounds on the graviton mass. The black horizontal line is a representative of the latter, corresponding to $m=10^{-32}$~eV.}
\label{fig:massVScoupling}
\end{figure}
%%%%%%%%%%
%

In this paper we show that bounds stronger than standard positivity constraints can be derived by taking into account the irreducible IR cross-sections under the dispersive integral, which are calculable within the EFT. 
In models where the forward amplitude is suppressed or the high-energy scattering is governed by soft dynamics (e.g. Galileons, massive gravity, dilatons, WZW-like theories \cite{Cheung:2016drk}),
as well as models with suppressed $2\rightarrow 2$ (but e.g.~enhanced $2\rightarrow 3$) amplitudes, our bounds are dramatically  stronger. 
These bounds can be used to place rigorous upper limits on the cutoff scale for certain EFT's or constrain the relevant couplings, in a way that is somewhat reminiscent of the  revived S-matrix bootstrap approach in four dimensions~\cite{Paulos:2017fhb}.
The procedure we use was originally suggested in~\cite{Nicolis:2009qm}, and later employed to estimate order-of-magnitude  bounds~\cite{Bellazzini:2016xrt,deRham:2017imi}; here we extend these arguments to sharp inequalities and bring this technique beyond amplitudes' positivity.

We discuss explicitly two relevant applications of the bounds: the EFT for a weakly broken Galileon \cite{Nicolis:2008in,Pirtskhalava:2015nla}, and the ghost-free massive gravity  theory \cite{deRham:2010ik,deRham:2010kj}, known also as dRGT massive gravity, or $\Lambda_3$-theory ($\Lambda_3$ is the strong coupling scale that remains in the Goldstone equivalence limit for the Galileon mode). 
Despite the encouraging recent results on the positivity conditions that ghost-free massive gravity must satisfy \cite{Cheung:2016yqr}, our constraints will provide a much stronger, and yet theoretically robust, \emph{lower bound} on the graviton mass $m$.
Indeed, our dispersion relations imply that the forward elastic amplitudes, which are \emph{suppressed} by $m$ at fixed $\Lambda_3$, must nevertheless be larger than a factor times the \emph{unsuppressed} elastic or inelastic cross-sections.
Resolving this tension requires a non-trivial lower bound for the graviton mass. 
Under the customarily accepted assumption that $\Lambda_3$ is the cutoff of the theory in  Minkowski background, i.e.~away from all massive sources, this lower bound reads $m\gtrsim$~100 keV  with $1\%$ uncertainty, which is grossly excluded observationally. 
Relaxing this assumptions by lowering the cutoff (i.e.~taking hierarchically separated values for the actual cutoff $\Lambda$ and the scale $\Lambda_3$ evaluated in Minkowski), we show that the dRGT massive gravity theory does not survive the combination of our bound with the experimental constraints on the graviton mass while being able to describe physical phenomena down to scales where gravity has been actually measured.
In other words, our result implies that the graviton mass can only be below the experimental upper bound at the expense of a premature break down of the theory (along with Vainshtein screening), therefore at the price of loosing predictivity at unacceptably large (macroscopic) distances.
This scale $\Lambda$ is where new physics states appear, which, importantly, is different than the scale where \emph{perturbative} unitarity would be lost \cite{Burrage:2012ja}, thus making our conclusions robust under our assumptions on the S-matrix.
{We anticipate these results in Fig.~\ref{fig:massVScoupling}: before this work all of the plane of graviton coupling and mass was theoretically allowed, as long as the parameters $c_3$ and $d_5$ of massive gravity satisfied the standard positivity constraints identified in Ref.~\cite{Cheung:2016yqr}, while now a point that falls in our excluded (gray) region means that the parameter space $(c_3, d_5)$ consistent with our bounds has shrank to an empty set, thus it is ruled out.}

In the following, we begin by deriving the new bounds in full generality, and then apply them to the Galileon theory, showing that Galileon-symmetry-breaking terms can not be arbitrarily small. 
This naturally leads us to ghost-free massive gravity, where we find the most dramatic implications of our bounds. 
Other relevant applications are discussed in the outlook.

%%%%%%%%%%%%%%%%%%%%%%%%%%%%%%%%%
\section{Dispersion Relations}

\begin{figure}[t]
\centering
\includegraphics[width=7cm]{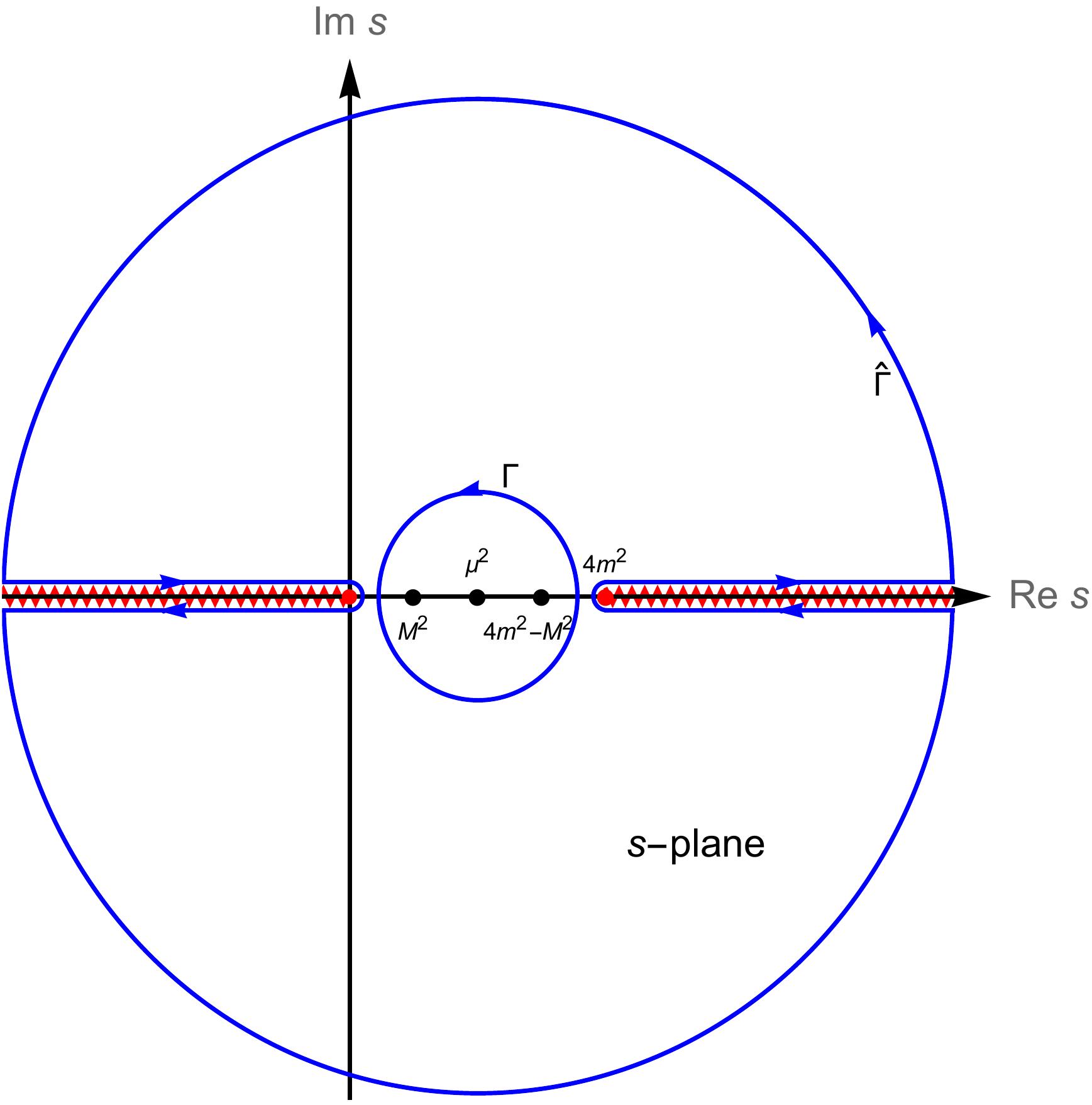}
\caption{Integration contours in the complex $s$-plane at fixed $t=0$, with poles at $s_1=M^2$ and $s_2=4m^2-M$. The point $
s=\mu^2$ is on the real axis between the branch-cuts shown in red. }
\label{fig:splane}
\end{figure}

Let us consider the center-of-mass  2-to-2 scattering amplitude $\M^{z_1z_2 z_3z _4}(s,t)$,  where the polarization functions are labeled  $z_i$. The Mandelstam variables\footnote{We use the mostly-plus Minkowski metric $(-,+,+,+)$, work with the relativistic normalization of one-particle states $\langle p, z|  p^\prime z^\prime \rangle=(2\pi)\delta^3(p-p^\prime)2E(p)\delta_{zz^\prime}$, and define the $\M$ operator from the S-matrix operator, $S=1+(2\pi)^4\delta^4(\sum k_i)i\M$.} are defined by  $s= -(k_1+k_2)^2$, $t=-(k_1+k_3)^2$, $u=-(k_1+k_4)^2$ and satisfy $s+t+u=4 m^2$, where $m$ is the mass of the scattered particles (all of the same species for ease of presentation).   
 Our arguments will require finite $m\neq 0$, yet they hold even for some massless theories (scalars, spin-1/2 fermions, and  softly broken $U(1)$ gauge theories), which have a smooth limit $m\rightarrow 0$ at least for the highest helicities, so that the bound can be derived  with an arbitrarily small but finite mass, before taking the massless limit.
We call,
\begin{equation}
\M^{z_1 z_2}(s)\equiv \M^{z_1 z_2 z_1 z_2}(s,t=0)\,,
\end{equation}
the forward elastic amplitude at $t=0$, and integrate $\M^{z_1 z_1}(s)/(s-\mu^2)^3$ along a closed contour $\Gamma$ in the complex $s$-plane, enclosing all the physical IR poles $s_i$ associated with the stable light degrees of freedom exchanged in the scattering (or its crossed-symmetric process), together with  the point $s=\mu^2$ lying on the real axis between $s=0$ and $s=4 m^2$,
\begin{equation}
\label{Sigmadef}
\Sigma_{\mathrm{IR}}^{z_1 z_2} \equiv \frac{1}{2\pi i} \oint_\Gamma ds\frac{\M^{z_1z_2}(s)}{(s-\mu^2)^3}\,, 
\end{equation}
see Fig.~\ref{fig:splane}.  
%
%%%%%%%%%%
%%%%%%%%%%
%
The $\Sigma_{\mathrm{IR}}^{z_1 z_2}$  is nothing but the sum of the IR residues,
\begin{equation}
\Sigma_{\mathrm{IR}}^{z_1 z_2}= \sum \underset{s=s_i,\mu^2}{\mathrm{Res}}\left[ \frac{\M^{
z_1 z_2}(s)}{(s-\mu^2)^3} \right] \,,
\end{equation}
and it is therefore calculable within the EFT.
Using Cauchy's integral theorem we deform the contour integral into  $\widehat{\Gamma}$ that runs just around the s-channel and u-channel branch-cuts, and goes along the big circle eventually sent to infinity.

The  polynomial in the denominator of Eq.~(\ref{Sigmadef}) has the lowest order that ensures the convergence of the dispersive integral in the UV, %\footnote{Whenever the tree-level EFT amplitude is dominated by $o(s^{n>2})$ at large $s$ and fixed $t=0$, it is convenient to use a higher polynomial $(s-\mu^2)^{n+1}$%, even though $(s-\mu^2)^3$ would suffice, because the resulting tree-level expressions for $\Sigma_{IR}$ are simpler and $\mu^2$ independent. Specifically, the right-most term in Eq.~(\ref{resinfty}) would get replaced by $1/n!(\partial^n\M)(\partial s^n)_{m^2 \ll s\ll \Lambda^2}$.}
a consequence of the Froissart-Martin asymptotic bound $|\M(s\rightarrow \infty)| < \mathrm{const}\,\cdot s\log^2 s$, which is always satisfied in any local massive QFT \cite{Froissart:1961ux,Martin:1965jj}. Thus $\lim_{s\rightarrow \infty}|\M(s)|/s^2\rightarrow 0 $, we can drop the boundary contribution and write $\Sigma_{IR}^{z_1 z_2}$ as an integral of the discontinuity $\mathrm{Disc}\M^{z_1 z_2}(s)\equiv \M^{z_1 z_2}(s+i\epsilon)-\M^{z_1 z_2}(s-i\epsilon)$ along the branch-cuts,
\begin{equation}
\Sigma_{\mathrm{IR}}^{z_1 z_2}= \frac{1}{2\pi i}\left(\int^{\infty}_{4m^2}ds   +\int^{0}_{-\infty}ds \right) \frac{\mathrm{Disc}\M^{z_1 z_2}(s)}{(s-\mu^2)^3}\,.
\end{equation}
The integral along the u-channel branch-cut  runs over non-physical values of $s=(-\infty,0)$, but can be expressed in terms of another physical amplitude, involving  anti-particles (identified by a bar over the spin label, i.e.~$\bar{z}$), and related to the former by crossing. Indeed, crossing particle $1$ and $3$ in the forward elastic limit $t=0$ implies, even for spinning particles ~\cite{Bellazzini:2016xrt}, that 
\begin{eqnarray}
\M^{z_1 z_2}(s)&=&\M^{-\bar{z}_1 z_2}(u=-s+4m^2)\quad\textrm{(helicity basis)} \,, \nonumber\\
\M^{z_1 z_2}(s)&=&\M^{\bar{z}_1 z_2}(u=-s+4m^2)\quad\textrm{(linear basis)} \,. \nonumber
\end{eqnarray}
We  will work in the helicity basis and recall when necessary that for $-\bar{z}\to\bar{z}$ we recover the results for linear polarizations. For particles that are their own antiparticles, $\bar{z}=z$. 

Finally, since amplitudes are real functions of complex variables, i.e.~$\M(s)^*=\M(s^*)$, the discontinuity above is proportional to the imaginary part, and one obtains the dispersion relation between IR and UV:
\begin{equation}
\label{eq:dispersive}
\Sigma_{\mathrm{IR}}^{z_1 z_2} \!=\!\!\int^{\infty}_{4m^2} 
\!\!\frac{ds}{\pi} \!\left(\frac{\mathrm{Im}\M^{z_1 z_2}(s)}{(s-\mu^2)^3}   +\frac{\mathrm{Im}\M^{-\bar{z}_1 z_2}(s)}{(s-4m^2+\mu^2)^3}\right).
\end{equation}

%%%%%%%%%%%%%%%%%%%%%%%%%%%%%%%%%
\section{Positivity and Beyond}

Unitarity of the $S$-matrix implies the optical theorem,
\begin{equation}
\label{optical}
\mathrm{Im}\M^{z_1 z_2}(s)=s \sqrt{1-4m^2/s}\cdot \sigma^{z_1 z_2}_{\mathrm{tot}}(s) >0\,,
\end{equation}
where $\sigma^{z_1z_2}_{\mathrm{tot}}(s)$ is the total cross-section $\sigma^{z_1 z_2}_{\mathrm{tot}}=\sum_X \sigma^{z_1 z_2\rightarrow X}$. 
Therefore the imaginary parts in the integrand Eq.~(\ref{eq:dispersive}) are strictly positive for any theory where particles 1 and 2 are interacting, as long as $0<\mu^2<4m^2$. One then obtains the rigorous positivity bound,
\begin{equation}
\label{eq:positivity1}
\Sigma_{\mathrm{IR}}^{z_1 z_2} >0\,.
\end{equation}
Since $\Sigma_{\mathrm{IR}}^{z_1 z_2}$ is calculable in the IR in terms of the Wilson coefficients,  Eq.~(\ref{eq:positivity1}) provides a non-trivial constraint on the EFT.

As a simple example consider the theory of a pseudo-Goldstone boson $\pi$, from an {approximate} global $U(1)$ symmetry which is broken spontaneously in the IR. The effective Lagrangian reads   
$\mathcal{L}_{\mathrm{EFT}}=-\frac{1}{2}(\partial\pi)^2+\frac{\lambda}{\Lambda^4}[(\partial\pi)^2+\ldots]^2- \epsilon^2\pi^2\left(\Lambda^2+c(\partial\pi)^2+\ldots \right)$,
where $\Lambda$ is the cutoff and $\lambda\sim o(1)$ (or even larger should the underlying dynamics be strongly coupled).
The parameters that break the approximate Goldstone shift symmetry $\pi\rightarrow \pi+\mathrm{const}$ are instead suppressed, naturally, by $\epsilon\ll 1$. From an EFT point of view, both signs of $\lambda$ are consistent with the symmetry; however $\Sigma_{\mathrm{IR}}=\lambda/2\Lambda^4$, so that only $\lambda>0$ is compatible with the positivity bound Eq.~(\ref{eq:positivity1}).
Unitary, local, causal and Lorentz invariant UV completions can generate only positive values for $\lambda$ in the IR~\cite{Adams:2006sv}. 
Notice that this statement does not depend on any finite value of the soft deformation $\epsilon$, which one is thus free to take arbitrarily small.

Like in the previous example,  $\Sigma_{\mathrm{IR}}$ is often calculable within the tree-level EFT, where the only discontinuities in the amplitude $\M^{\textrm{EFT}}$ are simple poles. In such a case we can use again Cauchy's theorem on the tree-level EFT amplitude so that  $\Sigma_{\mathrm{IR}}$ is more promptly calculated as minus the residue at infinity~\cite{Cheung:2016yqr}, 
\begin{equation}
\label{resinfty}
\Sigma_{\mathrm{IR}}^{z_1 z_2}= -\underset{s=\infty}{\mathrm{Res}} \left[ \frac{\M^{\textrm{EFT}}(s)}{(s-\mu^2)^3} \right]\,,
\end{equation}
up to small corrections. In addition, for amplitudes that scale as $\M^{\textrm{EFT}}(s) \sim s^2$ for large $s$ and  $t=0$ (as in e.g.~the Galileon or ghost-free massive gravity), we have $\Sigma_{\mathrm{IR}}^{z_1 z_2}={1 \over 2}(\partial^2\M^{\textrm{EFT}}/{\partial s^2})|_{m^2 \ll s}$.
In this case, the left-hand side of the dispersion relation Eq.~(\ref{eq:dispersive}) is $\mu^2$-independent and one can thus drop the dependence on $\mu^2$ of the right-hand side too.\\

% of order $g^2/(16\pi^2)(m^2/\Lambda^2)^n \ll 1$ from one-loop and higher-dimensional operators. These are usually tiny corrections, even for $g\sim 4\pi$, as long $m \ll \Lambda$ which is always the case in a relativistic EFT.

So far we invoked very general principles of QFT and derived positivity constraints on EFT's.
We can in fact extract more than positivity bounds by noticing that the total cross-section Eq.~(\ref{optical}) on the right-hand side of the dispersion relation Eq.~(\ref{eq:dispersive}) contains an irreducible contribution from IR physics, which is calculable within the EFT, by construction. The other contributions, e.g.~those from the UV, are incalculable with the EFT but are nevertheless always strictly positive, by unitarity.
Moreover, each final state $X$ in the total cross-section contributes positively too.
Therefore, an exact inequality follows from truncating the right-hand side of Eq.~(\ref{eq:dispersive}) at some energy $E^2\ll \Lambda^2$ below the cutoff $\Lambda$ of the EFT,
\begin{widetext}
\begin{equation}
\label{eq:dispersive2}
\Sigma_{\mathrm{IR}}^{z_1 z_2} > \sum_X
\int^{E^2}_{4m^2} 
\frac{ds}{\pi} \sqrt{1-4\frac{m^2}{s}}\left[\frac{s \sigma^{z_1 z_2\rightarrow X}(s) }{(s-\mu^2)^3}   +\frac{s \sigma^{-\bar{z}_1 z_2 \rightarrow X }(s) }{(s-4m^2+\mu^2)^3}\right]_{\mathrm{IR}} \,.
\end{equation}
\end{widetext}
Both sides are now calculable, hence the subscript $\mathrm{IR}$. The $\Sigma_{\mathrm{IR}}^{z_1 z_2}$ must not only be positive but \emph{strictly larger} than something which is itself positive and calculable within the EFT. 
Moreover, we can retain any subset $X$ of final states, independently on whether they are elastic or inelastic: the more channels and information are retained in the IR the more refined the resulting bound will be.

The information  provided by our bound Eq.~(\ref{eq:dispersive2}) is particularly interesting in theories where the elastic forward amplitude $\M^{z_1 z_2}$, which appears in the left-hand side, is parametrically suppressed compared to the non-forward or inelastic ones (that is  $\M^{z_1 z_2 z_1 z_2}(s,t\neq 0)$,  $\M^{z_1 z_2 z_3 z_4}(s,t)$, or more  generally $\M^{z_1 z_2 \rightarrow X}$), that appear in the right-hand side.
This tension results in constraints on the couplings and/or masses of the EFT, that include and go beyond the positivity of $\Sigma_{\mathrm{IR}}$.
For instance, Galileons have a suppressed forward amplitude: the leading term in the elastic amplitude, proportional to $stu$, 
actually vanishes at $t=0$ and  $\M^{z_1 z_2}$ is thus sensitive to the small Galileon-symmetry-breaking terms. On the other hand, neither the Galileon elastic cross-section nor the right-hand side of Eq.~(\ref{eq:dispersive2})  are suppressed.  
Massive gravity, the dilaton, WZW-like theories, as well as other models where $2\rightarrow 2$ is suppressed while $2\rightarrow 3$ is not, are other simple examples of theories that get non-trivial constraints from our bound Eq.~(\ref{eq:dispersive2}). 
Even in situations without  parametric suppression, our bound carries  important information:
it links elastic and inelastic cross-sections that might depend on different Wilson coefficients of the EFT.

Amplitudes in an EFT means finite, yet systematically improvable, accuracy $\delta$ in the calculation.
The main source of error for small masses is the truncation of the tower of higher-dimensional operators.  
Therefore, working to leading order (LO) in powers of $(E/\Lambda)^2$ and $(m/E)^2$ (hence also $(\mu/E)^2$), the bound Eq.~(\ref{eq:dispersive2}) takes a simpler form 
%\begin{widetext}
\begin{eqnarray}
\label{eq:LObound}
 \Sigma_{\mathrm{IR},\mathrm{LO}}^{z_1 z_2} & \!>\! & \sum_X
\int^{E^2} 
\!\!\!\frac{ds}{\pi s^2} \left[ \sigma^{z_1 z_2\rightarrow X}(s)  +\sigma^{z_1 -\bar{z}_2 \rightarrow X }(s) \right]_{\mathrm{IR},\mathrm{LO}}\nonumber\\ 
&&\times\left[1+o\left(\frac{m}{E}\right)^2+o\left(\frac{E}{\Lambda}\right)^2 \right] \,,
\end{eqnarray}
%\end{widetext}
where the error from the truncation
\begin{equation}
\label{errorestim}
o\left(\frac{E}{\Lambda}\right)^2=\left(c_{\mathrm{UV}}+o(1)\frac{ g_*^2}{16\pi^2}\ln\frac{E}{\Lambda}\right) \left(\frac{E}{\Lambda}\right)^2+\ldots
\end{equation}
is controlled by the (collective) coupling $g_*$ of the IR theory, which renormalizes the higher-dimensional operators that come with (unknown) Wilson coefficients  $c_{\mathrm{UV}}\sim o(1)$.\footnote{$c_{\mathrm{UV}}\gg 1$ would just signal the misidentification of what the actual LO hard-scattering amplitude is and would require including the operators with large $c_{\mathrm{UV}}$  within the LO amplitude.} 
The IR running effects, from $\Lambda$ to $E$, are an irreducible (yet improvable) source of error, whereas the UV contribution is  model dependent.

Choosing $E$ at or slightly below the cutoff $\Lambda$ gives just an order of magnitude estimate for the bound~\cite{Bellazzini:2016xrt,deRham:2017imi}. 
A rigorous bound can instead be obtained even for large couplings  $g_*\sim 4\pi$ and $c_{\mathrm{UV}}\sim 1$, by choosing a sufficiently small $(E/\Lambda)^2$. Percent accuracy can be achieved already with $E/\Lambda \approx 1/10$. Of course, nothing except more demanding calculations prevents us from reducing the error, e.g.~by working to all order in the mass or including next-to-next-to\ldots next-to-LO corrections, so that the truncation in the EFT expansion (or the running couplings) affects the result only by an even smaller relative error, $\mathrm{loops} \times o(E/\Lambda)^{n}$.\footnote{In addition, the LO may  possibly receive corrections from the logarithmic running of LO couplings. In the examples where our bounds are  interesting, symmetry are often at play and the LO operators do not actually get renormalized, except from small explicit breaking effects.}

%%%%%%%%%%%%%%%%%%%%%%%%%%%%%%%%%
\section{Galileon}

Let us consider the amplitude
\begin{equation}
\label{galamp}
\M(s,t)=g_*^2\left[-3\frac{stu}{\Lambda^6}+ \epsilon^2 \frac{s^2+t^2+u^2}{2\Lambda^4}+\ldots \right] \, ,
\end{equation}
for a single scalar $\pi$ whose $2\rightarrow 2$ hard-scattering limit is $o(s^3)$, whereas the forward scattering is $o(s^2)$ and suppressed by  $\epsilon^2\ll 1$.
The cutoff $\Lambda$ corresponds to a physical threshold for new states propagating on-shell, i.e.~the location of the first non-analyticity in the complex $s$-plane which is not accounted by loops of $\pi$.  We have factored out the overall coupling constant $g_*^2$ to make clear the distinction between the physical cutoff $\Lambda$ and other scales not associated to physical masses, such as decay constants, see Appendix~\ref{App:hbarcounting}. 

The amplitude Eq.~(\ref{galamp}) gives $\Sigma_{\mathrm{IR}}=g_*^2 \epsilon^2/\Lambda^4$ and $\sigma^{\pi\pi\rightarrow \pi\pi}=3g_*^4 s^5/(320\pi \Lambda^{12})+\ldots$~.\footnote{Curiously, there is a mild violation of the naive dimensional analysis (NDA) estimate $\epsilon^2_{\mathrm{NDA}}> \left(9g_*^2/16\pi^2\right)\left(E/\Lambda\right)^8$ \cite{Bellazzini:2016xrt} due to a $10\%$ cancellation in the phase-space integral $1/2 \int_{-1}^{1}d\cos\theta |stu|^2$, which returns $s^6(1/3+1/5-1/2)=s^6/30$ rather than $o(1)s^6$. } The bound Eq.~(\ref{eq:LObound}) reads in this case
\begin{equation}
\label{boundGalileon}
\epsilon^2 > \frac{3}{40} \left(\frac{g_*^2}{16\pi^2}\right)\left(\frac{E}{\Lambda}\right)^8\,,
\end{equation}
up to the relative error Eq.~(\ref{errorestim}).
The lesson to be learnt here is that $o(s^2)$ terms in the amplitude can not be too suppressed compared to the the $o(s^3)$ terms.
Choosing e.g.~a $30\%$ accuracy on the bound, corresponding to $\left(E/\Lambda\right)^8 \approx 10^{-2}$, one gets $\epsilon^2 >10^{-3} (1\pm 30\%)$ for a fully strongly coupled theory $g_*=4\pi$. Setting instead $E \sim \Lambda$ implies accepting $o(1)$  corrections to the bound $\epsilon^2 \gtrsim g_*^2/16\pi^2$.

The weakly broken Galileon Lagrangian \cite{Nicolis:2008in,Pirtskhalava:2015nla}, 
\begin{eqnarray}
\label{weakGalileon}
\mathcal{L}&=&-\frac{1}{2}(\partial_\mu \pi)^2 \bigg[1+\frac{c_3}{\Lambda^3}\square\pi+\frac{c_4}{\Lambda^6}\left((\square\pi)^2-\left(\partial_\mu\partial_\nu\pi\right)^2\right)\nonumber \\ && \qquad + c_5\left(\ldots\right) \bigg]
+ \frac{\lambda}{4\Lambda^4} \left[(\partial \pi)^2\right]^2-\frac{m^2}{2}\pi^2 \, ,
\end{eqnarray}
has suppressed symmetry-breaking terms $\lambda\ll c_3^2, c_4$ and $m^2\ll \Lambda^2$.  It reproduces the scattering amplitude Eq.~(\ref{galamp}) with the identification
\begin{equation}
\label{galmatch}
c_3^2-2c_4 =4g_*^2\,,\qquad  \frac{\lambda}{\Lambda^4}+\frac{c_3^2 m^2}{2\Lambda^6}=\frac{g_*^2 \epsilon^2}{\Lambda^4}=\Sigma_{\mathrm{IR}}\,.
\end{equation}
In the massless limit, or more generally for $c_3^2 m^2/\Lambda^2 \ll \lambda$ (a natural hierarchy given that $\lambda$ preserves a shift symmetry while $m^2$ does not), the bound Eq.~(\ref{boundGalileon}) shows not only that $\lambda$ must be positive, but (parametrically) at most one-loop factor away from $(c_3^2-2c_4)/4$,
\begin{equation}
\label{eq:exgal1}
\lambda >  \frac{3}{640}\frac{\left(c_3^2-2c_4\right)^2}{16\pi^2}\left(\frac{E}{\Lambda}\right)^8\,.
\end{equation}
For a massive Galileon with negligible $\lambda$ and $c_3 \neq 0$, one gets a lower bound on the mass  ,  
\begin{equation}
\label{eq:exgal2}
m^2> \Lambda^2 \left(\frac{3}{320}\right)\frac{\left(c_3-2c_4/c_3\right)^2}{16\pi^2}\left(\frac{E}{\Lambda}\right)^8 \,,
\end{equation} 
where $\left(E/\Lambda\right)^8 \approx 10^{-2}$ for a $30\%$ accuracy. Therefore, the Galileon-symmetry-breaking terms can not be arbitrarily suppressed.

Our analysis has been performed at tree level, but the results Eqs.~(\ref{eq:exgal1}, \ref{eq:exgal2}) hold when loop effects are included. For instance, the $2 \to 2$ amplitude receives a contribution from a one-loop diagram with only $c_3$ insertions that scales as $(s/\Lambda^{2})^6 c_3^4/16\pi^2$, possibly with a $\log$. The correction to $\Sigma_{\mathrm{IR}}$ goes instead like $(m/\Lambda)^6 c_3^4 m^2 /16\pi^2 \Lambda^6$, with a real $\log$ since $\Sigma_{\mathrm{IR}}$ is evaluated for $\mu^2$ below threshold. Therefore, this contribution is negligible and consistent with our bound Eq.~(\ref{eq:exgal2}) as long as $m \ll E \ll \Lambda$. As one expects for such a higher derivative theory as the Galileon, {unsuppressed loops affect higher-dimensional operators only, and correspond to nothing but next-to-LO corrections to both sides of the inequality.}
Similarly, the contribution to the dispersive integral  from the symmetry-breaking $(\partial \pi)^4$ interaction is negligible as long as $\lambda/c_3^2 \ll (E/\Lambda)^2$, which is consistent with our bound Eq.~(\ref{eq:exgal1}) again as long as $E \ll \Lambda$.

%%%%%%%%%%%%%%%%%%%%%%%%%%%%%%%%%
\section{Massive Gravity}

The previous bounds on Galileons are unfortunately not directly applicable to models of modified gravity, which contain other IR degrees of freedom affecting $\Sigma_{\mathrm{IR}}$ significantly, such as e.g.~a massless graviton like in Horndeski theories~\cite{Koyama:2013paa}. In that case both sides of the inequality would be ill-defined at the Coulomb singularity $t=0$, because of the massless spin-2 state exchanged in the $t$-channel. Alternative ideas or extra assumptions are needed to deal with a massless graviton, see e.g.~Ref.'s.~\cite{Camanho:2014apa,Bellazzini:2015cra,Cheung:2016wjt,Benakli:2015qlh}.

In a massive gravity theory the situation is instead more favourable, as a finite graviton mass plays a double role: it regulates the IR singularity and tips the $o(s^2)$ term (vanishing in the forward and decoupling limit) to either positive or negative values depending on the parameters of the theory, which get therefore constrained by the positivity of $\Sigma_{\mathrm{IR}}$ \cite{Cheung:2016yqr}. Notice that one can not directly interpret the results obtained above for the scalar Galileon as the longitudinal component of the massive graviton, since the IR dynamics is different and we are after next-to-decoupling effects {(i.e.~$\sim m^2$)} in $\Sigma_{\mathrm{IR}}$: for example, in the scattering of the Galileon scalar mode, the helicity-2 mode exchanged in the $t$-channel gives a contribution to the amplitude that is as large as the contribution from the exchange of the scalar mode. 

The action for ghost-free massive gravity \cite{deRham:2010ik,deRham:2010kj} is given by (for reviews see Ref.'s~\cite{Hinterbichler:2011tt,deRham:2014zqa}), 
\begin{equation}
S=\int d^4 x \sqrt{-g}\left[ \frac{\mp^2}{2} R - \frac{\mp^2 m^2}{8}V(g,h) \right] \,,
\end{equation}
where $\mp=(8\pi G)^{-1/2}$ is the reduced Planck mass,  $g_{\mu\nu}=\eta_{\mu\nu}+h_{\mu\nu}$ is an effective metric written in term of the Minkowski metric $\eta_{\mu\nu}$ (with mostly $+$ signature) and a spin-2 graviton field $h_{\mu\nu}$ in the unitary gauge, $R$ is the Ricci scalar for $g_{\mu\nu}$, and $V(g,h)=V_{2}+V_3+V_4$ is the soft graviton potential,
\begin{align}
V_2(g,h)= & b_1 \langle h^2 \rangle+ b_2\langle h \rangle^2 \,, \\ 
V_3(g,h)= & c_1 \langle h^3 \rangle+ c_2\langle h^2 \rangle \langle h \rangle+ c_3\langle h \rangle^3 \,, \\ 
V_4(g,h)= & d_1\langle h^4 \rangle+ d_2\langle h^3 \rangle \langle h \rangle+ d_3\langle h^2 \rangle^2 \\
\nonumber &+ d_4 \langle h^2 \rangle  \langle h \rangle^2 +d_5 \langle h \rangle^4 \,,
\end{align}
with $\langle h\rangle\equiv h_{\mu\nu}g^{\mu\nu}$, $\langle h^2\rangle\equiv g^{\mu\nu}h_{\nu\rho}g^{\rho\sigma}h_{\sigma\mu}$, etc. The coefficients depend on just two parameters, $c_3$ and $d_5$, after imposing the ghost-free conditions
\begin{align}
b_1= 1 \,,\, & \,\, b_2=-1 \,, \\
c_1=2c_3+\frac{1}{2}\,,\,  & \,\,  c_2=-3c_3-\frac{1}{2}\,, \\
d_1=-6d_5+\frac{3}{2} c_3+\frac{5}{16} \,,\, & \,\, d_2=8 d_5-\frac{3}{2} c_3-\frac{1}{4} \,, \\
d_3=3d_5-\frac{3}{4} c_3-\frac{1}{16}\,,\, & \,\, d_4=-6d_5+\frac{3}{4}c_3\,.  
\end{align}

Since the graviton is its own antiparticle, it is convenient to work with linear polarizations, since they make the crossed amplitudes, and in turn the bound, neater \cite{Bellazzini:2016xrt,Cheung:2016yqr,Bellazzini:2015cra}. For example, the LO bound with linear polarizations reads 
\begin{equation}
\label{eq:LOboundlinear}
 \Sigma_{\mathrm{IR},\mathrm{LO}}^{z_1 z_2} > \sum_X
\frac{2}{\pi}\int^{E^2} 
\frac{ds}{s^2} \left[ \sigma^{z_1 z_2\rightarrow X}(s) \right]_{\mathrm{IR},\mathrm{LO}} \,.
\end{equation}
Adopting the basis of polarizations reported in Appendix~\ref{App:pol}, we have two tensor polarizations ($T$, $T^\prime$) that do not grow with energy, two vector polarizations ($V$, $V^\prime$) that grow linearly with energy, and one scalar polarization  ($S$) that grows quadratically with the energy.
 
We calculate the amplitudes for different initial and final state configurations and find that $\Sigma^{z_1 z_2}_{\mathrm{IR}} \sim m^2/\Lambda_3^6$ is suppressed by the small graviton mass, where \cite{ArkaniHamed:2002sp}
\begin{equation}
\Lambda_3 \equiv (m^2 \mp)^{1/3}\,.
\end{equation}
On the other hand, we find that the cross-sections are not generically suppressed by $m$: hence, a small mass is at odds with our  bound Eq.~(\ref{eq:LOboundlinear}). Resolving this tension results in non-trivial constraints on the theory \cite{Bellazzini:2016xrt}, beyond the positivity bounds derived in Ref.~\cite{Cheung:2016yqr}. 

The amplitudes for  $SS,V^{(\prime)}V^{(\prime)},V^{(\prime)}S$ elastic scatterings have the following suppressed residues,
\begin{align}
& \Sigma^{SS}_{\mathrm{IR}} = \frac{2m^2}{9\Lambda_3^6}\left(7-6c_3(1+3c_3)+48 d_5\right)>0 \, , \nonumber \\
& \Sigma^{VV}_{\mathrm{IR}} = \Sigma^{V'V'}_{\mathrm{IR}} \! = \frac{m^2}{16 \Lambda_3^6} \left(5+72 c_3-240 c_3^2\right) >0 \label{sresidues} \, , 
\end{align}
\begin{align}
& \Sigma^{VV^\prime}_{\mathrm{IR}} = \frac{m^2}{16  \Lambda_3^6} \left(23-72 c_3+144 c_3^2+192 d_5\right)>0 \, , \nonumber \\
& \Sigma^{VS}_{\mathrm{IR}}= \Sigma^{V^\prime S}_{\mathrm{IR}}= \frac{m^2}{48 \Lambda_3^6} \left(91-312c_3+432 c_3^2+384 d_5\right)>0 \, . \nonumber 
\end{align}
In contrast, the hard-scattering limits of the amplitudes that enter the right-hand side of Eq.~(\ref{eq:LOboundlinear}) are unsuppressed. For $s,t \gg m^2$ these read,
\begin{align}
&\M^{SSSS}= \frac{s t (s+t)}{6\Lambda_3^6}\left( 1-4c_3(1-9c_3)+64 d_5\right) \, , \nonumber \\
&\M^{VVVV}= \M^{V'V'V'V'}= \frac{9st(s+t) }{32\Lambda_3^6} (1-4c_3)^2 \label{highexsec} \, , \nonumber  \\
&\M^{VV^\prime VV^\prime}=  \frac{3t^3 }{32 \Lambda_3^6 }(1-4c_3)^2 \, ,  \\
&\M^{VSVS}=  \frac{3t}{4\Lambda_3^6}\Big(c_3(1-2c_3)(s^2+st-t^2)\nonumber\\ & \hspace{4cm} -\frac{5s^2+5st-9t^2}{72}\Big) \, ,\nonumber \\
&\M^{V^\prime S V^\prime S}= \frac{1}{96 \Lambda_3^6}\big( st(s+t)(7 - 24 c_3 + 432 c_3^2 + 768 d_5)\nonumber\\ & \hspace{4cm}  -9t(1 - 4 c3)^2 t^2\big) \, . \nonumber
\end{align}
It is convenient to recall also the bound 
\begin{equation}
\label{yellowregionupper}
\frac{m^2}{36\Lambda_3^6}\left(35 + 60 c_3 - 468 c_3^2 - 192 d_5\right)>0 \, ,
\end{equation}
which follows from the positivity of the residue of maximally-mixed $ST$ polarizations, i.e.~$\Sigma_{\mathrm{IR}}^{TT}+\Sigma_{\mathrm{IR}}^{SS}+2\Sigma_{\mathrm{IR}}^{TSTS}+4\Sigma_{\mathrm{IR}}^{TTSS}>0$%(with a slight abuse of notation)
, where the expressions for these $\Sigma_{\mathrm{IR}}$ are given in Appendix~\ref{App:nonelastic}. 

At this point we choose the energy scale $E$ in Eq.~(\ref{eq:LOboundlinear}) below the cutoff, $E \ll \Lambda$, so that the EFT calculation of the cross-sections is trustworthy, and above the mass $E \gg m$, so that the EFT hard-scattering amplitudes Eq.~(\ref{highexsec}) are dominating such cross-sections. We define
\begin{equation}
\delta\equiv \left(\frac{E}{\Lambda}\right)^2 \, , 
%= \frac{1}{g_*} \left(\frac{E}{\Lambda_3 }\right)^3\,,\qquad g_* \equiv  \left(\frac{\Lambda}{\Lambda_3}\right)^3
\end{equation}
that controls the accuracy of the EFT calculation, and obtain
\begin{equation}
\label{boundmassivegr1}
F_i(c_3,d_5)>  \left(\frac{4\pi \mp}{m}\right) \left(\frac{g_*}{4\pi}\right)^4 \delta^6\,, 
\end{equation}
where we have defined 
\begin{equation}
 \label{coupdef}
g_* \equiv  \left(\frac{\Lambda}{\Lambda_3}\right)^3\,.
\end{equation}
The functions $F_i(c_3,d_5)$ are given by
\begin{widetext}
\begin{align}
& F_{SS} = \left[960\frac{7-6c_3(1+3c_3)+48 d_5}{\left(1-4c_3(1-9c_3)+64 d_5\right)^2}\right]^{3/2} \, , \nonumber\\ 
& F_{VV} = \left[\left(\frac{2560}{27}\right) \frac{5+72 c_3-240 c_3^2}{(1-4c_3)^4} \right]^{3/2} \label{ffunctions} \, , \\
& F_{VV^\prime} = \left[ \left(\frac{896}{9}\right)\frac{23-72c_3+144 c_3^2+192 d_5}{(1-4c_3)^4}\right]^{3/2} \, , \nonumber\\
& F_{VS} =  \! \left[\frac{80640 \left(91-312c_3+432 c_3^2+384 d_5\right)}{1975-29808c_3 (1-2c_3)(1-4c_3+8c_3^2)}\right]^{3/2} \, ,\nonumber\\
& F_{V^\prime S} =  \left[\frac{80640 \left(91-312c_3+432 c_3^2+384 d_5\right)}{1891 - 21504 d_5 + 48 (c_3 (-649 + 6 c_3 (649 + 24 c_3 (-41 + 153 c_3))) + 10752 c_3 (1 + 6 c_3) d_5 + 86016 d_5^2)}\right]^{3/2} \nonumber \,.
\end{align}
\end{widetext}
The five inequalities following from Eq.~(\ref{boundmassivegr1}) are the main result of this section: they imply \emph{lower bounds} on the graviton mass, which  for a fixed $g_*$ can not be arbitrarily small compared to  $4\pi \mp$ (this, incidentally, is the largest cutoff for quantum gravity). As remarked in Ref.~\cite{Bellazzini:2016xrt}, one can take $m\rightarrow 0$ only by sending $g_*\rightarrow 0$ as well. These bounds represent a much improved, sharper and more conservative version of the rough estimate presented in Ref.~\cite{Bellazzini:2016xrt}. As we discuss below, $g_*$ cannot be taken arbitrarily small either without seriously compromising the predictive power of the EFT for massive gravity.

\subsection*{Implications}

The bounds Eq.~(\ref{boundmassivegr1}) can be read in several ways: as constraints on the plane of the graviton potential parameters $(c_3,d_5)$ for a given graviton mass $m$ and ratio $(\Lambda/\Lambda_3)^3=g_*$, as a constraint on $g_*$ for fixed $m$ at a given point in the $(c_3,d_5)$ region allowed by positivity, or equivalently as a bound on the graviton mass for fixed coupling at that point. For these last two interpretations, an \emph{absolute} constraint on $g_*$ versus $m$ can be derived.

We begin with a discussion of the bounds on the parameters $c_3$ and $d_5$ inside $F_i$.
The experimental \emph{upper limit} on the graviton mass is extremely stringent, $m\lesssim 10^{-32}-10^{-30}$~eV, depending on the type of experiment and theory assumptions behind it (see Ref.~\cite{deRham:2016nuf} for a critical discussion).
\begin{figure*}[t]
\begin{center}
\includegraphics[width=16.5cm]{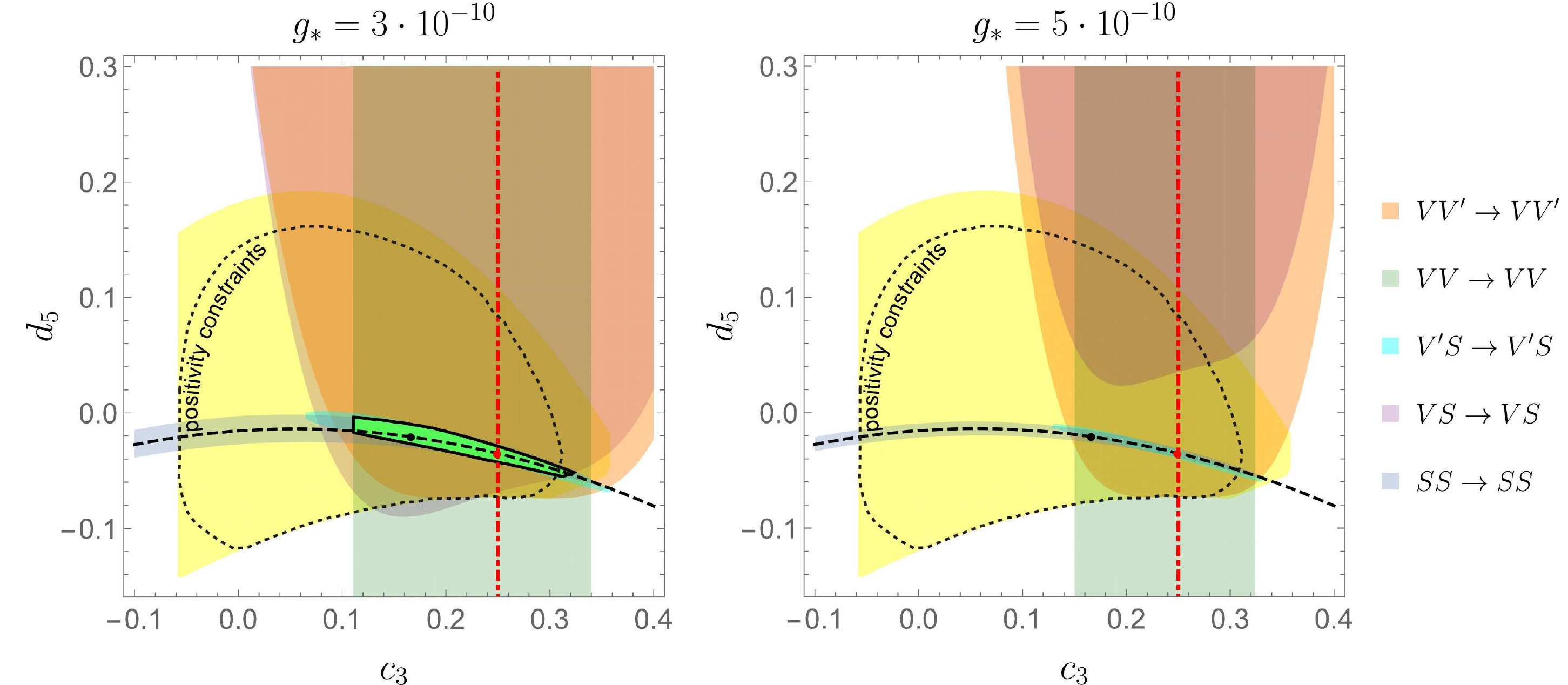}
\caption{Exclusion plots in the $(c_3,d_5)$ plane for ghost-free massive gravity, for fixed accuracy $\delta= 1\%$, mass $m=10^{-32}$~eV, and coupling $g_* = 3 \, (5) \cdot 10^{-10}$ in the left (right) panel. The two plots illustrate how the region allowed by our bounds (green region inside the solid line) shrinks to the point of disappearing as the coupling is increased above $4.5\cdot 10^{-10}$.
The yellow region is allowed by the standard positivity constraints, Eqs.~(\ref{sresidues}, \ref{yellowregionupper}), whose optimized version from Ref.~\cite{Cheung:2016yqr} is delimited by the dotted black line.
The other regions are the ones consistent with our new bounds, Eq.~(\ref{boundmassivegr1}), the different colors corresponding to each of the $F_i$ in Eq.~(\ref{ffunctions}), as specified in the legend. 
 On the dash-dotted red (dashed black) line, $F_{VV}$ ($F_{SS}$) vanishes, and so it does the corresponding bound.  On the red dot $(c_3, d_5)=(1/4,-9/256)$ the vector and scalar modes decouple from the tensors, but not from each other, and on the black dot $(c_3, d_5)=(1/6,-1/48)$ the scalar mode decouples from the tensor mode and itself.}
\label{fig:MGbound}
\end{center}
\end{figure*} 
Taking $m=10^{-32}$~eV as benchmark, we show in Fig.~\ref{fig:MGbound} the constraints on $c_3$ and $d_5$, for given $g_*$; the colored regions being allowed by our constraints. The yellow region is determined from the standard positivity constraints Eqs.~(\ref{sresidues}, \ref{yellowregionupper}),  while the others follow from our new bounds in Eq.~(\ref{boundmassivegr1}). Already for $g_* = 5 \cdot 10^{-10}$ (right panel), corresponding to the situation where $\Lambda$ and $\Lambda_3$ are about a factor $10^3$ away from each other, our bounds do not admit any solution in the $(c_3,d_5)$ plane, likewise for any larger value of $g_*$ at the same mass.
Comparing both panels of Fig.~\ref{fig:MGbound}, we note that as $g_*$ increases the constraints from $F_{VV}$ or $F_{VV^\prime}$ alone single out essentially a narrow band around the line $c_3=1/4$, in agreement with the causality arguments of Ref.'s~\cite{Camanho:2016opx,HinterbiCHLer:2017qyt}. Similarly, the constraint from $F_{SS}$ converges quickly to the line $1-4c_3(1-9c_3)+64 d_5=0$ {(while $F_{V'S}$ restricts such a line to a range of $c_3$ values)}. The intersection point (red dot in Fig.~\ref{fig:MGbound}), $(c_3,d_5)=(1/4,-9/256)$, is finally removed by $F_{VS}$. 
In substance, the intersection region in the right panel of Fig.~\ref{fig:MGbound} is empty.
Instead, a small island (colored in green and delimited by a solid black line) survives in the left panel, which corresponds to a smaller $g_* = 3 \cdot 10^{-10}$. 

To find the absolute maximum value of $g_*$ below which our bounds allow for a solution, or, analogously, the minimum value of $m$, we write Eq.~(\ref{boundmassivegr1}) as
\begin{equation}
\label{massbound1}
m  > 1.2\cdot 10^{12} \, \mathrm{eV}\, \left(\frac{g_*}{1}\right)^4 \left(\frac{\delta}{1\%}\right)^6 \frac{1}{F_i(c_3,d_5)} \,,
\end{equation}
and note that at each point $(c_3, d_5)$, the bound is determined by the smallest $F_i$.
Therefore, the maximum of the (continuous) function $\mathrm{min}\{F_i\}(c_3, d_5)$ in the positivity region sets the most conservative bound. This corresponds to $(\hat{c}_3,\hat{d}_5) \approx (0.18, -0.017)$ (close to the black point in Fig.~\ref{fig:MGbound}) for which $F_{VS} \approx 4.6 \cdot 10^6$, yielding the lower bound 
\begin{equation}
\label{FVSmaxrelation}
m > 10^{-32} \, \mathrm{eV} \left(\frac{g_*}{4.5 \cdot 10^{-10}}\right)^4  \left(\frac{\delta}{1\%}\right)^6\,.
\end{equation} 
We recall that the direct experimental constraint on the graviton mass is $m \lesssim 10^{-32}$~eV,
implying that any value $g_* \gtrsim  4.5 \cdot 10^{-10}$ is excluded, irrespectively of the values of $(c_3,d_5)$.
This situation is summarized in Fig.~\ref{fig:massVScoupling}: for values of $(g_*,m)$ that fall in the gray region, the allowed island of $(c_3,d_5)$ parameter space has completely disappeared.
Besides, even slightly stronger bounds can be obtained by working with the non-elastic channels (see Appendix~\ref{App:nonelastic}), while if we were to admit a slightly larger uncertainty, e.g.~$\delta = 5\%$, the upper bound on $g_*$ would increase by one order of magnitude.\\

At this point the crucial question is: What is the physical meaning of $g_*$, the relation between the physical cutoff $\Lambda$ and the scale $\Lambda_3$? Can the UV completion  be arbitrarily weakly coupled $g_*\lesssim 10^{-10}$ \cite{Bellazzini:2016xrt}? To our knowledge, most literature of massive gravity  has so far either identified the cutoff $\Lambda$ with the scale $\Lambda_3$, or assumed $\Lambda\gg\Lambda_3$, so that one would expect $g_* \gtrsim 1$. These values are now grossly excluded by our bounds.

What about hierarchical values for $\Lambda$ and $\Lambda_3$ corresponding to tiny values for $g_*$? 
From a theoretical point of view, $\Lambda$ and $\Lambda_3$ scale differently with $\hbar$,  
so that their ratio actually changes when units are changed, in such a way that $g_*$ indeed scales like a coupling constant (see  Appendix~\ref{App:hbarcounting}).    
This is fully analogous to the difference between a vacuum expectation value $v$ (VEV) and the mass of a particle $\sim coupling \times v$, e.g.~the $W$-boson mass $m_W \sim g v$.
The crucial point then is that the cutoff $\Lambda$ is a physical scale, which differs from $\Lambda_3$ that instead does not have the right dimension to represent a cutoff.
Since $\Lambda_3^{-1} \approx 320 \, \mathrm{km}\, \left(m/10^{-32} \, \mathrm{eV} \right)^{-2/3}$, a 
very small coupling $g_*$ translates into a very low cutoff (large in units of distance)
\begin{equation}
\label{barecutoff}
\!\!\! \! \Lambda \simeq \left(4.1 \cdot 10^5 \,\mathrm{km}\right)^{-1} \left(\frac{g_*}{4.5 \cdot 10^{-10}}\right)^{1/3} \left(\frac{m}{10^{-32} \, \mathrm{eV}}\right)^{2/3} \!\!\!\!\!\!\! .
\end{equation}
This is clearly problematic, a major drawback of the theory of massive gravity once we recall that general relativity (GR) has been precisely tested at much smaller distances, down to the mm or even below, see e.g.~Ref.'s~\cite{Tan:2016vwu, Kapner:2006si,deRham:2016nuf}. 
In other words, while GR, taken as an EFT, has been experimentally shown to have a cutoff below the mm, thus providing a good description of gravitational phenomena from (sub)millimeter to cosmological distances, in the light of our bounds dRGT fails to describe the same phenomena below scales of the order of the Earth-Moon distance.
\\

More specifically, let us consider the experimental tests of massive gravity in the form of bounds on fifth forces from the precise measurements of the Earth-Moon precession $\delta \phi$ \cite{Dvali:2002vf,Williams:2004qba,deRham:2014zqa}.
Due to the Vainshtein screening \cite{Vainshtein:1972sx,Deffayet:2001uk}, which is generically dominated by the Galileon cubic interactions in the $(c_3,d_5)$ region allowed by our bounds, the force mediated by the scalar mode compared to the standard gravitational one is $F_S/F_{\textrm{GR}} \sim (r/r_V)^{3/2}$, where $r_V = (M/4\pi \mp)^{1/3} \Lambda_3^{-1} = (M/4\pi m^2 \mp^2)^{1/3}$ is the Vainshtein radius associated with the (static and spherically symmetric) source under consideration, in this case the Earth, $M = M_\oplus$.
Before our bound, one would find that at lunar distances $r=r_{\oplus L} \approx 3.8 \cdot 10^5$~km, the ratio of forces and thus also the precession $\delta \phi \sim \pi (F_S/F_{\textrm{GR}})$, even if very small for $m=10^{-32}$~eV, would be borderline compatible with the very high accuracy of present measurements $\sim 10^{-11}$.
Now our bound Eq.~(\ref{barecutoff}) shows that the EFT is not valid already for $r \sim 1/\Lambda > r_{\oplus L}$. 
This implies that the Vainshtein screening should receive important corrections before reaching the (inverse) cutoff $1/\Lambda$ and, moreover, it means that new degrees of freedom should become active at that scale: two effects that likely impair the fifth-force suppression and hinder the agreement with the precise measurement of the Earth-Moon precession.\footnote{Even extremely weakly coupled new degrees of freedom can give $o(1)$ deviations from the non-analytic dependence over the couplings when a new state goes on-shell. A simple example is the exchange of a new weakly coupled particle at threshold, which gives maximal phase shift in the amplitude regardless of the size of the overall coupling.} 

Besides, one should note that the cutoff in Eq.~(\ref{barecutoff}) holds in Minkowski space and not necessarily in regions near massive bodies, such as the Earth, where classical non-linearities are important and Vainshtein screening is active.
In such non-trivial backgrounds, the strong coupling scale $\Lambda_3$ gets redressed as $\Lambda_3 \to z \Lambda_3$, with $z \gg 1$ deep inside the Vainshtein region \cite{Nicolis:2004qq}.
However, this Vainshtein rescaling (or redressing, not to be confused with the Vainshtein screening) relies on the assumption that the tower of effective operators is such that only the building blocks of the type $\partial\partial\pi/\Lambda_3^3$ are unsuppressed and dominate (we work here for simplicity with the Stueckelberg mode $\pi$ in the decoupling limit), and therefore it does not \emph{generically} extend to operators suppressed by extra derivatives, $(\partial/\Lambda)^n$, sensitive to the bona fide cutoff of the EFT. The cutoff for the fluctuations keeps being $\Lambda$, which, following our bound Eq.~(\ref{barecutoff}), is encountered much before the strong coupling scale, i.e.~$\Lambda \ll \Lambda_3$. 
In this sense, and unless ad-hoc assumptions are made, such a Vainshtein rescaling of the cutoff could be relevant to extend the EFT validity only for $\Lambda \gg \Lambda_3$ (or $g_*\gg 1$), but this is exactly the region ruled out by our bounds.\\

The tension between the bounds Eqs.~(\ref{massbound1}, \ref{FVSmaxrelation}), direct limits on the graviton mass, and fifth-force experiments, leads us to conclude that ghost-free massive gravity is not a proper contender of GR for describing gravitational phenomena, in that the EFT can not tell e.g.~whether an apple would fall to the ground from the tree, float mid-air or else go up.\footnote{Our conclusion is general and does not depend on special tunings of the potential parameters within the positivity region. Indeed, for e.g.~$d_5 = -c_3/8$, where Vainshtein screening is essentially that of the quartic Galileon \cite{Chkareuli:2011te,Berezhiani:2013dw} instead of the cubic Galileon, the experimental upper bound on the graviton mass is $m = 10^{-30}$~eV, corresponding to a cutoff that %, while allowing for describing the Earth-Moon system, 
is still very large $1/\Lambda \gtrsim \mbox{few} \cdot 10^4$~km (of the order of the geostationary orbit of satellites).}
This constitutes a major concern for the theory of massive gravity in view of our bounds, which warrants extending the theory in the ``UV'' in such a way to describe the relevant gravitational phenomena while remaining consistent with experimental tests (i.e.~the new gravitational dynamics remaining undetected) not only in lunar experiments but also down to the mm.
In fact, ``UV''  corresponds here to macroscopic distances, of order $\mbox{few} \cdot 10^5$~km.
To emphasize this fact, we can speculate about non-generic situations (i.e.~departing from NDA expectations, most likely requiring fine-tuning) for what regards the EFT expansion.
One (trivial) possibility is that all Wilson coefficients associated with the operators containing extra derivatives happen to be suppressed, which entails the validity (and thus calculability) of the theory extends beyond $\Lambda$, even in flat space. In this case we can effectively choose $E$ in Eq.~(\ref{eq:LOboundlinear}) larger than $\Lambda$, and therefore our bounds get stronger as well, so that the theory would still be ruled out.
Alternatively, we can imagine that the whole tower of operators associated with the extra derivative terms come with the right powers of fields (and coefficients) in order for the true cutoff $\Lambda$ (as well as $\Lambda_3$) to be raised, i.e.~Vainshtein redressed, in a (certain, appropriately chosen) non-trivial background, but not necessarily extending the calculability in Minkowski space.
In this case, rescaling $\Lambda$ at the Earth's surface, $r=r_\oplus$ (thus assuming a spherical background), one arrives at
\begin{align}
\label{VainCutoff}
\Lambda_\oplus & \sim \left( \frac{r_V}{r_{\oplus}} \right)^{3/4} \Lambda \\
%= \left(\frac{M_\oplus}{4\pi \mp} \frac{1}{(r_{\oplus}\Lambda_3)^3}\right)^{1/4}  \Lambda
&\approx (37\,\mathrm{m})^{-1}  \left(\frac{g_*}{4.5 \cdot 10^{-10}}\right)^{1/3}\left(\frac{m}{10^{-32} \, \mathrm{eV}}\right)^{1/6} \,,
\nonumber
\end{align} 
Even with this extra epicycle, the redressed cutoff of massive gravity is still orders of magnitudes larger than the (sub)millimeter scale, where GR has been successfully tested. This fact is illustrated in Fig.~\ref{fig:massVScoupling}. We also note that more aggressive bounds can be derived by accepting large uncertainties, e.g.~for $\delta = 10\%$ then $\Lambda_\oplus \approx (120\,\mathrm{m})^{-1}$.\\

In summary, our theoretical bounds either rule out massive gravity or show that the theory is unable to make predictions at scales where GR instead does and in agreement with experimental observations. This last observation calls for new ideas on extending the theory in the UV. Of course, violation of the assumptions that led to our bounds (e.g. Lorentz invariance, polynomial boundedness) is also a logical possibility, although not much different from finding explicit UV completions, since also requires non-trivial dynamics in the UV. Finally, note that considering either smaller couplings or masses (e.g.~$m\sim H_0\sim 10^{-33}$~eV to explain cosmic acceleration) only aggravates the problem, since the (inverse) cutoff is increased.

%%%%%%%%%%%%%%%%%%%%%%%%%%%%%%%%%
\section{Outlook}

Positivity bounds  are statements that arise from first principles such unitarity, analyticity, and crossing symmetry of the Lorentz invariant S-matrix. They have proven to be very useful because they set non-perturbative theoretical constraints even in strongly coupled theories, giving information that goes well beyond the mere use of symmetries. In this paper we went beyond positivity bounds and derived rigorous inequalities for amplitudes that are calculable in the IR via an  EFT approach.  The dispersive integral in the IR is not only positive but also calculable, with an error from truncating the EFT towers of higher-dimensional operators that can be tamed thanks to separation of scales,  which is what makes the EFT useful in the first place.
 
Our results, while simple and general, can be applied straightforwardly to several EFT's. The {implementation on} interesting theories such as the weakly-broken Galileon and the ghost-free massive gravity that we explored in this paper are extremely rewarding.

Taken at face value, our bounds rule out dRGT massive gravity in a large range of masses $m$ and couplings $g_* = (\Lambda/\Lambda_3)^3$. In fact, our constraints on the EFT are qualitatively different from previous bounds, in that they crucially incorporate $g_*$, which controls the size of the allowed island of parameter space $(c_3,d_5)$ of ghost-free massive gravity. Furthermore, when combined with the experimental constraints on the graviton mass, our bounds seriously limit the realm of predictivity of massive gravity, since the physical cutoff $\Lambda$ is forced well below $\Lambda_3 = (m^2 \mp)^{1/3}$ (specifically $g_* \lesssim 4.5\times 10^{-10}$ with $1\%$ uncertainty for $m = 10^{-32}$ eV), leaving an EFT that does not stand competition with GR already below macroscopically large distances: of the order of the Earth-Moon distance (without extra non-generic assumptions about the tower of effective operators), or in the 50 to 100 meter range (if Vainshtein redressing the cutoff). Below these scales, the EFT is \emph{not even wrong}. It would certainly be compelling to find UV completions in order to assess if the theory is able to pass experimental constraints at those scales.

Needless to say, our bounds neither apply to Lorentz-violating models of massive gravity (e.g.~\cite{Blas:2014ira}), nor to theories with a massless graviton: one can avoid our bounds by dropping any of the assumptions on the S-matrix that led to them.

There are several directions where our bounds can find fruitful applications. The most immediate ideas involve  theories with Goldstone particles, e.g.~the EFT for the Goldstino from SUSY breaking or the R-axion from R-symmety breaking, and the dilaton from scale-symmetry breaking. In these theories there exist universal couplings that are set by the various decay constants, and there are also non-universal parameters whose sizes and signs are often not accessible with the standard positivity bounds. Our results would allow instead to relate these non-universal parameters to the decay constants and extract thus non-trivial information on the EFT's, of phenomenological relevance, see e.g.~\cite{Bellazzini:2017neg,Bruggisser:2016ixa,Bellazzini:2017bkb,Bellazzini:2013fga,Bellazzini:2012vz}.
Another phenomenologically interesting direction would be towards theories that have suppressed 2-to-2 amplitudes but unsuppressed 2-to-3 amplitudes, as those discussed e.g.~in \cite{Cheung:2016drk}.

It is also attractive to recast our bounds in diverse spacetime dimensions. We tested the consistency of the conjectured a-theorem in $d=6$ (see e.g.~\cite{Elvang:2012st}) with our bounds, at least when the RG flow is initiated by spontaneous breaking of scale invariance. In this case, we anticipate here that for large coefficients of the Weyl- and diffeomorphism-invariant 4-derivative term $b$, the variation of the a-anomaly $\Delta a$ can not be negative without violating our bound. More specifically, the (conventionally chosen dimensionless) points in the plane $(b,\Delta a)$ must fall in a band, parametrically of the form $b > \mbox{loop} \times (\frac{3}{2}\Delta a- b^2)^2>0$, which implies only a finite range $0<b<b_*$ consistent with a negative $\Delta a$ and our bound. Considering instead lower dimensional spacetimes, one could investigate what our bound implies e.g.~for massive gravity theories in $d=3$ \cite{Bergshoeff:2009hq,Deser:1981wh}. 

One further stimulating avenue is to use our bounds to extend the no-go theorems for massless higher spin particles in flat space (see e.g.~\cite{Weinberg:1995mt,Weinberg:1964ew,Grisaru:1976vm,Porrati:2008rm,Porrati:2012rd}) to the case of small but finite masses. 
While the no-go theorems can be evaded with arbitrarily small masses, we expect that our bound can, analogously to the case we explored for massive gravity, put a limit on how light higher-spin particles can be relative to the cutoff of the theory. Such a result would represent a quantitative assessment of why light higher-spin particles can not emerge, even in principle, in non-gravitational theories without sending the cutoff to zero or making them  decouple. 

One important open question, that for the time being remains elusive, is whether it is possible (at least under extra assumptions) to extend our results to theories with massless particles and with spin $J\geq 2$. If that would be the case, the resulting bounds would provide new insights on the long-distance universal properties of the UV completion of quantum gravity, such as string theory.  The bounds would also apply to IR modifications of GR  such as  Horndeski-like theories,  where the graviton remains massless.

%%%%%%%%%%%%%%%%%%%%%%%%%%%%%%%%%

\medskip

%%%%%%%%%%%%%%%%%%%%%%%%%%%%%%%%%
\subsection*{Acknowledgements}
We thank Matt Lewandowski, Matthew McCullough, David Pirtskhalava, Riccardo Rattazzi, Andrew Tolley, Enrico Trincherini, and Filippo Vernizzi for useful discussions. We thank Duccio Pappadopulo, Massimo Porrati and Gabriele Trevisan for useful comments. 
We thank Claudia de Rham, Scott Melville, Andrew Tolley, and Shuangyong Zhou for correspondence concerning the Vainshtein mechanism.
B.B.~thanks Cliff Cheung and Grant Remmen for correspondence.
B.B.~is supported in part by the MIUR-FIRB grant RBFR12H1MW ``A New Strong Force, the origin of masses and the LHC''; B.B.~thanks Marco Cirelli and the LPTHE for the kind hospitality during the completion of this work, and Roberto Contino and Enrico Trincherini for the kind hospitality at the SNS. J.S.~and B.B.~would like to express a special thanks to the Mainz Institute for Theoretical Physics (MITP) for its hospitality and support.

%%%%%%%%%%%%%%%%%%%%%%%%%%%%%%%%%
\appendix

%%%%%%%%%%%%%%%%%%%%%%%%%%%%%%%%%
\section{$g_*$-counting via $\hbar$-counting}
\label{App:hbarcounting}

In this appendix we recall how dimensional analysis is useful to extract the scaling with respect to coupling constants. 

Rescaling the units from $\hbar=1$ to $\hbar\neq 1$ while keeping $c=1$ reintroduces a conversion factor between energy (or momentum) units, $\E$, and length (or time) units, $\ell$, i.e.~$\ell=\hbar/\E$. 
With canonically normalized kinetic terms, we have the following scaling with $\hbar$:
$[A]=\E [\hbar]^{-1/2}$,  $[\partial]=\E [\hbar]^{-1}$, $[m]=\E$, and $[g_*]= [\hbar]^{-1/2}$,  where $g_*$ is (a collective name for) coupling constant(s) and $m$ a physical mass. 
%By dimensional analysis, the mass enters in the form of (inverse) Compton wave-length $\lambdabar=\hbar/m$ in the Lagrangian density, schematically as $\mathcal{L} \sim -(\partial A)^2-m^2/\hbar^2 A^2$.    
%This scaling is promptly checked for example in a non-abelian gauge theory in the Higgs phase.  
%One can also check it against the hamiltonian $H=\frac{1}{2m}\left(\mathbf{p}- g_* \mathbf{A}\right)^2+g_* A^0$ for a particle in an external electromagnetic field in quantum mechanics. 
Note for instance that a Higgs quartic coupling $\lambda$ scales really like a coupling squared $[\lambda]=[g_*^2]$.
Quantum corrections scale indeed like powers of the dimensionless quantity  $g_*^2\hbar/(16\pi^2)$ or $\lambda \hbar/(16\pi^2)$, so that they are important for $g^2_* \sim 16\pi^2/\hbar\sim \lambda$, as long as there are no large dimensionless number (such as e.g.~the number of species). Extending this dimensional analysis to fermions, it is immediate to see that Yukawa couplings scale also like $\hbar^{-1/2}$. 

Importantly, the relation between VEV's, couplings, physical masses and the associated  Compton lengths  is
\begin{equation}
\label{vevsandcoup}
[\lambdabar^{-1}]=\left[\frac{m}{\hbar}\right]=[g_* \langle A\rangle] \,.
\end{equation}
Therefore a coupling times a VEV is nothing but an inverse physical length, which can be converted to a physical mass by plugging in the conversion factor, aka $\hbar$. In other words, the appearance of a coupling in Eq.~(\ref{vevsandcoup}) tells us that parametrically   {\it VEV's are to masses (or Compton lengths) like apples are to oranges}.\footnote{We thank Riccardo Rattazzi who inspired this adage, with his interventions at the J. Hopkins workshop in Budapest in 2017.}  
The immediate consequence of this exercise is that the reduced Planck mass $\mp$ has units of a VEV, $[\mp]=[A]$, and not of a physical mass scale, in full analogy with an axion decay constant $[f_a]=[A]$. The UV completion of GR should enter at some physical energy $g_* \mp \hbar$, which is parametrically different than $\mp$ %, even after setting $\hbar=1$, 
because of the coupling $g_*$. 
 
This analysis with $\hbar\neq 1$ is useful to keep track of the appropriate $g_*$ counting; 
the structure of a generic Lagrangian that automatically reproduces it is,
\begin{equation}
\label{NDAlagra}
\mathcal{L}=\frac{\Lambda^4}{g_*^2}\widehat{\mathcal{L}}\left(\frac{\partial}{\Lambda}, \frac{g_* A}{\Lambda},  \frac{g_*\psi}{\Lambda^{3/2}}\right) \, ,
\end{equation}
where $\Lambda$ is a physical mass scale and $\widehat{\mathcal{L}}$ is a polynomial with dimensionless coefficients, where we have restored $\hbar=1$ units. The Lagrangian Eq.~(\ref{NDAlagra}) accounts for the intuitive fact that any field insertion in a given non-trivial process  requires including a coupling constant as well. A class of simple theories with only one coupling and one scale (see e.g.~Ref.~\cite{Giudice:2007fh}) are those where all dimensionless coefficients in $\widehat{\mathcal{L}}$ are of the same order (except for those associated with terms that break a symmetry, which can be naturally suppressed). This structure represents a generalization of the naive counting of factors of $4\pi$, routinely used in strongly coupled EFT's in particle physics (see e.g.~\cite{Cohen:1997rt}), which goes under the name of naive dimensional analysis (NDA).

With the $g_*$-counting at hand, we immediately recognize that the (strong coupling) scale $\Lambda_3^3 = m^2 \mp$ conventionally used in massive gravity is not parametrically a physical  threshold, since it misses a coupling constant. This is made manifest by the fact that the graviton mass is a physical mass scale but $\mp$ is only a VEV.  
Alternatively, in the decoupling limit the coefficient of the cubic Galileon must carry a coupling $g_*$, that is $[c_3]=[g_*]$ to match the general scaling of Eq.~(\ref{NDAlagra}). The actual correct parametric scaling for the physical cutoff is thus  $\Lambda^3= g_* \Lambda_3^3$. A weakly coupled theory corresponds to a suppressed $\Lambda$ relative to $\Lambda_3$, i.e.~$g_* \ll 1$, like a weakly coupled UV completion of GR corresponds to states entering much earlier than $4\pi \mp$. 

%%%%%%%%%%%%%%%%%%%%%%%%%%%%%%%%%
\section{Polarizations}
\label{App:pol}

We adopt the following basis of linear polarizations 
\begin{align}
& \left(\epsilon^{T}(\mathbf{k}_1)\right)^{\mu\nu} =\frac{1}{\sqrt{2}}\left( \begin{array}{cccc}
0 & 0 & 0 & 0\\
0 & 1 & 0 & 0\\
0 & 0 & -1 & 0\\
0 & 0 & 0 & 0
\end{array}\right)^{\mu\nu}\,, \nonumber  \\
&\left(\epsilon^{T^\prime}(\mathbf{k}_1)\right)^{\mu\nu} =\frac{1}{\sqrt{2}}\left( \begin{array}{cccc}
0 & 0 & 0 & 0\\
0 & 0 & 1 & 0\\
0 & 1 & 0 & 0\\
0 & 0 & 0 & 0
\end{array}\right)^{\mu\nu}\,, \label{polk1} \\
&\left( \epsilon^{V}(\mathbf{k}_1)\right)^{\mu\nu} =\frac{1}{\sqrt{2} m }\left( \begin{array}{cccc}
0 & k^z_1 & 0 & 0\\
 k^z_1 & 0 & 0 & E\\
0 & 0 & 0 & 0\\
0 & E & 0 & 0
\end{array}\right)^{\mu\nu}\,, \nonumber  \\
&\left(\epsilon^{V^\prime}(\mathbf{k}_1)\right)^{\mu\nu} =\frac{1}{\sqrt{2} m }\left( \begin{array}{cccc}
0 & 0& k^z_1  & 0\\
0 & 0 & 0 & 0\\
k^z_1  & 0 & 0 & E\\
0 & 0& E & 0
\end{array}\right)^{\mu\nu}\,, \nonumber \\
&\left(\epsilon^{S}(\mathbf{k}_1)\right)^{\mu\nu} =\sqrt{\frac{2}{3}}\left( \begin{array}{cccc}
\dfrac{k^{z\,2}_1}{m^2} & 0& 0  & \dfrac{k^z_1 E}{m^2}\\
0 & -1/2 & 0 & 0\\
0 & 0 & -1/2 & 0\\
\dfrac{k^z_1 E}{m^2} & 0& 0 & \dfrac{E^2}{m^2}
\end{array}\right)^{\mu\nu} \,, \nonumber
\end{align}
which are associated to the particle $k^\mu_1=(E,\mathbf{k}_1)=(E_1,0,0,k_1^z)$ moving along the $z$-axis with $E^2=\mathbf{k}_1^2+m^2$. These polarizations are real, symmetric, traceless, orthogonal, transverse to $k_1$, and with norm $\epsilon^*_{\mu\nu}\epsilon^{\nu\mu}=1$.\footnote{We are taking the same matrix entries of Ref.~\cite{Cheung:2016yqr}, except that that we have removed the $i$ factor from the vector polarizations and taken all upper Lorentz indexes. We checked that our choice satisfies the completeness relation. The $i$ factor is never important in elastic amplitudes, but it should actually be included whenever considering mixed-helicity states that include vector components, as done in \cite{Cheung:2016yqr}.}
The polarizations associated to the other momenta $k_{i}^\mu$ in the 2-to-2 scattering, in the center of mass frame, are obtained by a Lorentz transformation of those in Eq.~(\ref{polk1}), for instance
\begin{equation}
\left(\epsilon^{V}(\mathbf{k}_3)\right)^{\mu\nu}=\tensor{R}{^\mu_{\mu^\prime}} \tensor{R}{^\nu_{\nu^\prime}}  \left(\epsilon^{V}(\mathbf{k}_1)\right)^{\mu^\prime\nu^\prime}
\end{equation} 
with $\tensor{R}{^\mu_{\mu^\prime}}$ the rotation around the $y$-axis by $\cos\theta=1+2t/(s-4m^2)$ such that $k_3=R \, k_1$. While this definition is valid and legitimate, it corresponds effectively to consider $k_1$ as the canonical reference vector, rather than $(m,0,0,0)$, %^T
upon which constructing the massive one-particle states via boosting. Alternatively, it means that the standard Lorentz transformation that sends $(m,0,0,0)$ %^T
{to $k_1$} is a boost along the $z$-axis followed by a rotation that sends $\hat{z}$ to {$\hat{k_1}$} {(like it is done e.g.~for massless particle in Ref.~\cite{Weinberg:1995mt})}, rather than the sequence rotation-boost-rotation usually adopted for massive states \cite{Weinberg:1995mt}. The advantage of our convention is that it removes the little group matrix that would otherwise act on the polarization indexes $z=T,T^\prime, V,V^\prime, S$ when performing the rotations that send $k_1$ to $k_i$ (the Wigner rotation must be adapted accordingly too). For massless particles the differences between the two conventions is essentially immaterial as the little group acts just like phases.  

%%%%%%%%%%%%%%%%%%%%%%%%%%%%%%%%%
\section{Non-elastic channels in massive gravity}
\label{App:nonelastic}

%
%%%%%%%%%%
\begin{figure}[t]
\begin{center}
\includegraphics[width=8.5cm]{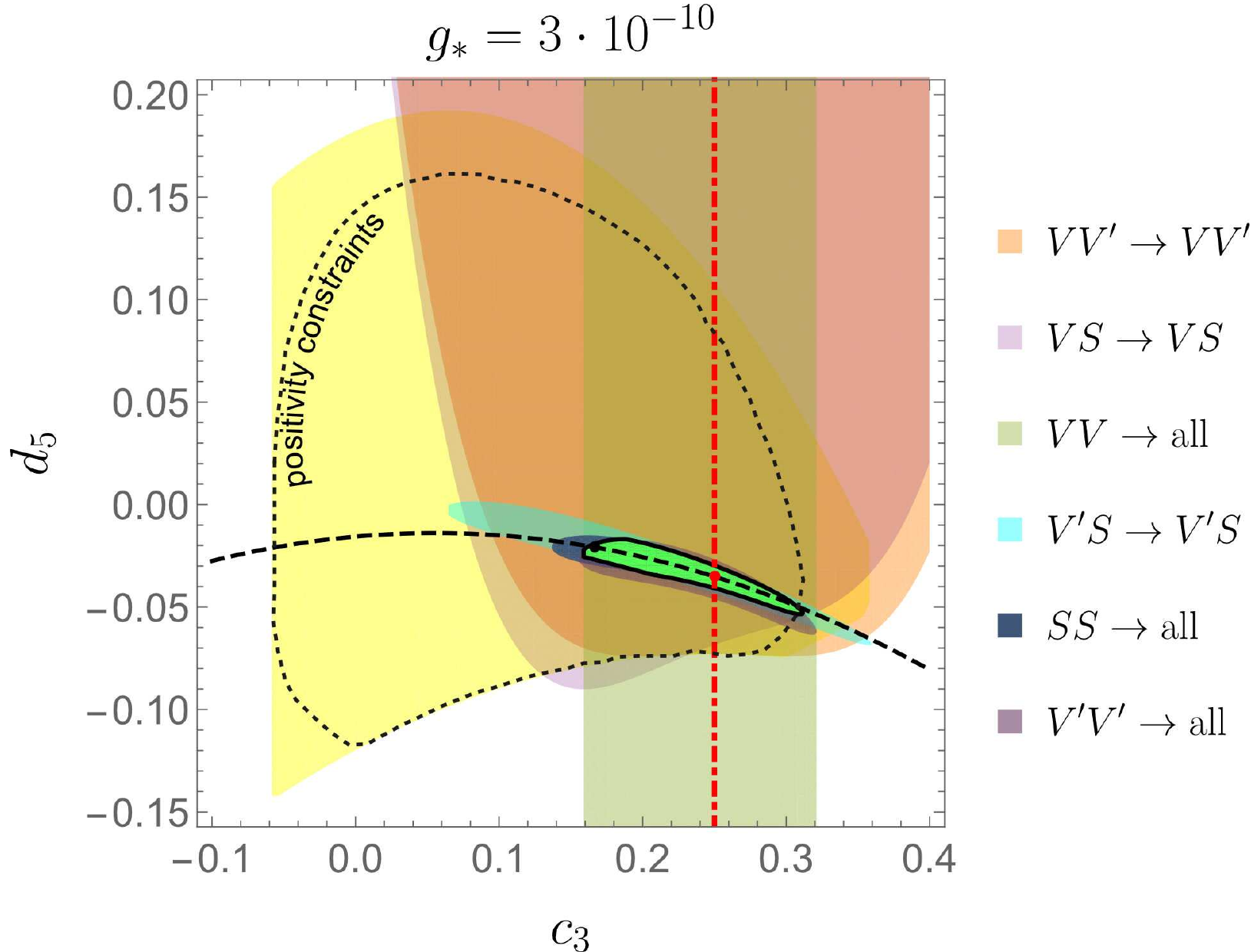}
\caption{Exclusion plots in the $(c_3,d_5)$ plane for ghost-free massive gravity, for fixed accuracy $\delta= 1\%$, mass $m=10^{-32}$~eV, and coupling $g_* = 3 \cdot 10^{-10}$, using inelastic channels. See the caption of Fig.~\ref{fig:MGbound} for other information about the figure. For couplings larger than $g_* \approx 4.4 \cdot 10^{-10}$ the green island disappears, for the same value of the mass, and the model is ruled out.}
\label{fig:inelasticplot}
\end{center}
\end{figure}
%%%%%%%%%%
%

We report in this appendix the impact of the inelastic channels in setting the lower bound on the graviton mass using Eq.~(\ref{eq:LOboundlinear}) with $X\neq z_1 z_2$. 
Their effect is not very significant, see Fig.~\ref{fig:inelasticplot}.  
The resulting maximum value of the new $\mathrm{min}\{F_i\}(c_3, d_5)$ function, that includes now the inelastic channels, is $4.3 \cdot 10^{6}$, at the point $(\hat{c}_3,\hat{d}_5) \approx (0.19, -0.022)$. This lightly lower value barely improves the bound in Eq.~(\ref{FVSmaxrelation}), obtained with the elastic channels only.

The inelastic cross-sections on the right-hand side of Eq.~(\ref{eq:LOboundlinear}) are calculated using the hard-scattering amplitudes for $s,t\gg m^2$
\begin{align}
\label{eq:inelastichannels}
\M^{VVSS}(s,t) = & \M^{SSVV}(s,t) =- \M^{VSVS}(t,s) \,, \\ 
\M^{V^\prime V^\prime SS}(s,t) = & \M^{SS V^\prime V^\prime }(s,t) =  - \M^{V^\prime SV^\prime S}(t,s) \,, \nonumber \\
\M^{VVV^\prime V^\prime}(s,t)  = & \M^{V^\prime V^\prime VV}(s,t) =  \M^{VV^\prime VV^\prime}(t,s)\,.  \nonumber
\end{align}
They are related to the elastic amplitudes simply by exchanging $s\leftrightarrow t$, up to an overall sign (which is not physical as it can be changed by redefining the phases of the polarizations, e.g.~adding a factor $i$ to the $V$ and $V^\prime$ polarizations). 
Note that this relation is a manifestation of crossing symmetry in the Goldstone-equivalence limit.
Amplitudes involving tensor polarizations scale more slowly with energy, as $s/\mp^2=s m^4/\Lambda_3^6$ or $s^2/(m^2 \mp^2)=s^2 m^2/\Lambda_3^6$, in the hard-scattering limit $s, t\gg m^2$, for instance
\begin{align}
\!\! \M^{TTTT} & =-\frac{(s^2 + s t + t^2)^2}{\mp^2 s t (s + t)}+\frac{9(1-4c_3^2)t(s+t)}{2\mp^2 s }\,.
\end{align}
Therefore they are not useful to derive bounds with our methods. Moreover, crossing symmetry, relating e.g.~the hard-scattering limits of $\M^{TSTS}$ and $\M^{TTSS}$, is not just exchanging $s\leftrightarrow t$, even in the decoupling limit, precisely because one is sensitive in this case to the subleading corrections. 

For completeness, we report here also the following residues
\begin{align}
&\Sigma_{\mathrm{IR}}^{TT}= \frac{m^2}{\Lambda_3^6} \,, \qquad  \Sigma_{\mathrm{IR}}^{TSTS} = \frac{m^2}{\Lambda_3^6} (5-12 c_3) \,, \\
&\Sigma_{\mathrm{IR}}^{TTSS} = -\frac{m^2}{2 \Lambda_3^6} (1 - 8 c_3 + 24 c_3^2 + 16 d_5) \,, \nonumber 
\end{align}
which have been used in the main text  to obtain the bound Eq.~(\ref{yellowregionupper}) from maximally mixed $ST$ states \cite{Cheung:2016yqr}.

\medskip

%\bibliography{xxx}

\end{document}